\documentclass{moriond}

\def\Journal#1#2#3#4{{#1} {\bf #2}, #3 (#4)}

\def\PLB{{\em Phys. Lett.}  B}

\def\PRD{{\em Phys. Rev.} D}

\def\be{\begin{equation}}
\def\ee{\end{equation}}
\def\bea{\begin{eqnarray}}
\def\eea{\end{eqnarray}}

\usepackage{upgreek}
\usepackage{lineno}

\newcommand{\Photo}{}

\begin{document}
\vspace*{4cm}
\title{Observation of the decay $\mathrm{Z}\to\psi\,\ell^+\ell^-\to\mu^+\mu^-\ell^+\ell^-$ with the CMS detector}

\author{S. Leontsinis, \\ on behalf of the CMS collaboration}

\address{University of Colorado Boulder,\\
Department of Physics, Duane Physics E1B32,\\
2000 Colorado Ave, Boulder, CO 80309-0390,
USA}

\maketitle\abstracts{
The observation of the $\mathrm{Z}$ boson rare decay to a $\psi$ meson and two oppositely charged same-flavour leptons, $\ell^+ \ell^-$, where $\psi$ represents the sum of $\mathrm{J}/\psi$ and $\psi(\mathrm{2S})\to\mathrm{J}/\psi\, X$, and $\ell=\mu,\mathrm{e}$, is presented. The data sample used corresponds to an integrated luminosity of $35.9\,\mathrm{fb}^{-1}$ of proton-proton collisions at a center-of-mass energy of $13\,\mathrm{TeV}$ accumulated by the CMS experiment at the LHC. The signal is observed with a significance in excess of 5 standard deviations. Removing contributions from $\psi(\mathrm{2S})$ decays to $\mathrm{J}/\psi$, the signal is interpreted as being entirely from $\mathrm{Z}\to\mathrm{J}/\psi\,\ell^+\ell^-$, with its fiducial branching fraction relative to that of the decay $\mathrm{Z}\to\mu^+\mu^-\mu^+\mu^-$ measured to be $$\frac{\mathcal{B}(\mathrm{Z}\to\mathrm{J}/\psi\,\ell^+\ell^-)}{\mathcal{B}(\mathrm{Z}\to\mu^+\mu^-\mu^+\mu^-)}=0.70\pm 0.18\, \mathrm{(stat)} \pm 0.05\, \mathrm{(syst)}.$$ This result is obtained with the assumption of no $\mathrm{J}/\psi$ polarisation, where extreme polarisation scenarios can create $-24\%$ to $+22\%$ variations.
}

\section{Introduction}

\begin{figure}[h!]
\centering
{\includegraphics[width=.35\textwidth]{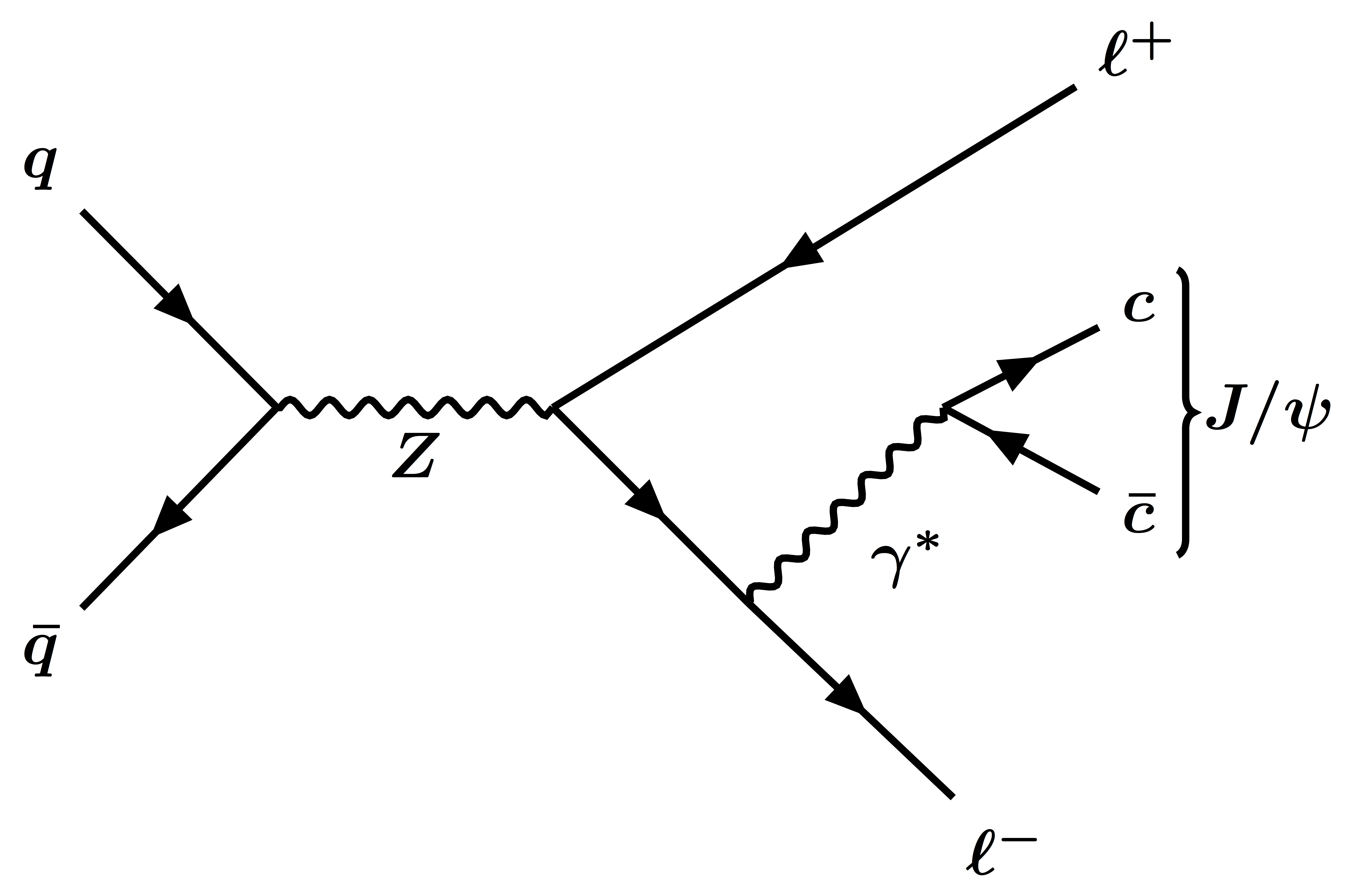}}
{\includegraphics[width=.3\textwidth]{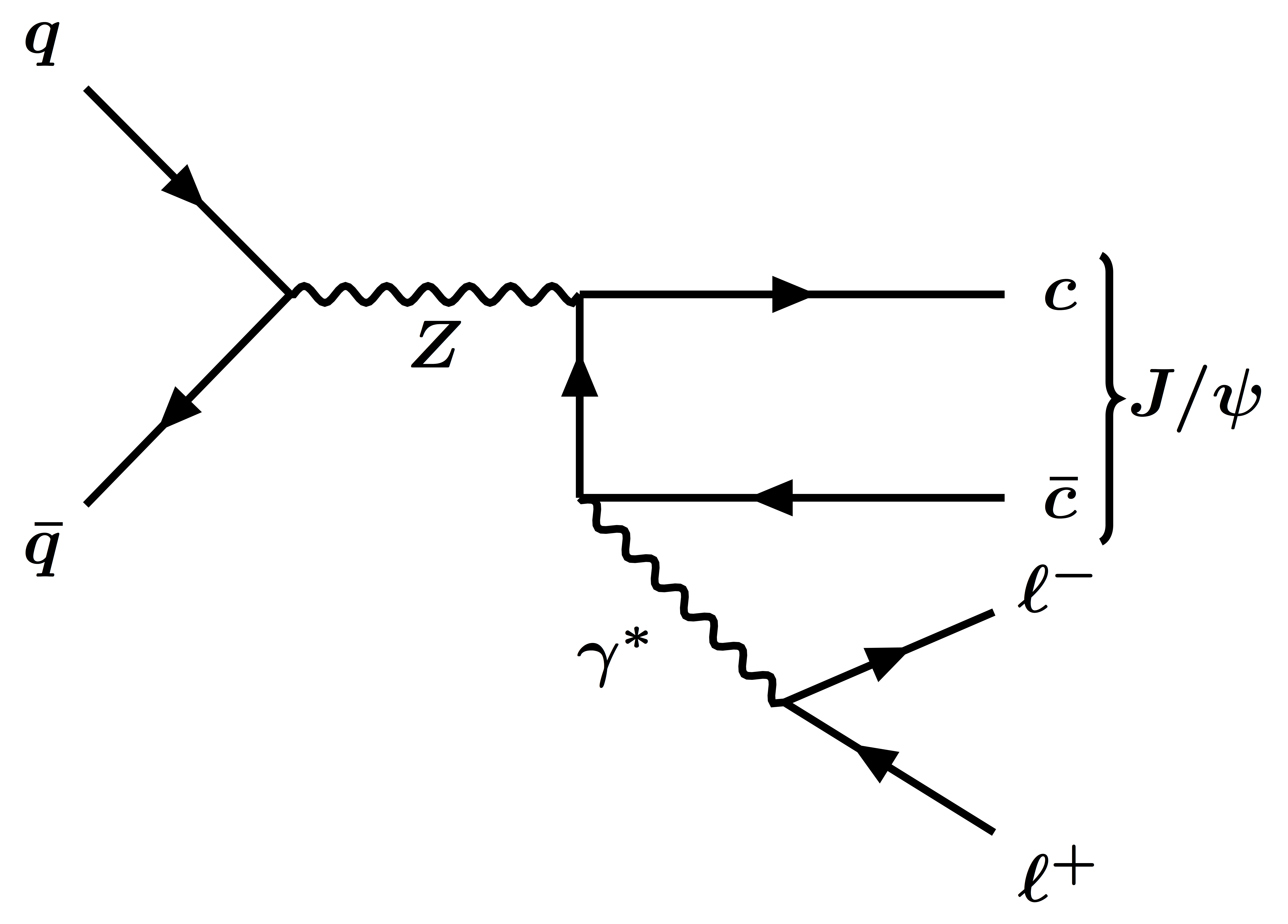}}
\caption{Leading-order diagrams for the $\mathrm{Z}\to\mathrm{J}/\psi\ell^+\ell^-$ process. The diagram on the left is the dominant production mechanism$^{5}$.}
\label{fig:feynman}
\end{figure}

The amazing performance of the LHC~\cite{LHCmachine} provides CMS with a large sample of $\mathrm{Z}$ bosons. With such a large amount of data, the CMS collaboration can now probe rare decay channels, such as $\mathrm{Z}\to\mathrm{V}\ell^+\ell^-$, where $\mathrm{V}$ is a vector meson. Such decays where consider by various theory groups~\cite{theorypaper1,theorypaper2,theorypaper3} during the LEP era. Theoretical calculations of this process, illustrated in figure~\ref{fig:feynman} for the case where $\mathrm{V}=\mathrm{J}/\psi$ predict a branching fraction of $(6.7$--$7.7)\times 10^{-7}$.

Although there are many searches for the $\mathrm{Z}$ boson to decay in the $\mathrm{J}/\psi\gamma$ final state~\cite{ATLAS1,ATLAS2}, the $\mathrm{J}/\psi\ell^+\ell^-$ decay mode provides a cleaner experimental signature and larger branching fraction~\cite{theorypaper3}. This is because the lepton propagator in the diagram shown in figure~\ref{fig:feynman} left is of the order of $1/M^2_{\mathrm{J}/\psi}$, while on the one on the right is of the order of $1/M^2_\mathrm{Z}$. This boosts the diagram on figure~\ref{fig:feynman} left by a factor $M^2_\mathrm{Z}/M^2_{\mathrm{J}/\psi}$ making it the dominant $\mathrm{J}/\psi$ meson production mechanism in electromagnetic $\mathrm{Z}$ decays. 

In this article, the first observation of $\mathrm{Z}\to\psi\ell^+\ell^-$ is presented~\cite{mypaper}, where $\psi$ sums over $\mathrm{J}/\psi$ and $\psi(\mathrm{2S})\to\mathrm{J}/\psi\,X$, and $\ell=\mu,\mathrm{e}$. The data sample used is $35.9\,\mathrm{fb}^{-1}$ of proton-proton collisions recorded by the CMS experiment~\cite{CMS} at the LHC at $\sqrt{s}=13\,\mathrm{TeV}$ center-of-mass energy. After removing contributions from $\psi\mathrm{(2S)}\to\mathrm{J}/\psi\, X$, the branching fraction ratio ($\mathcal{R}_{\mathrm{J}/\psi\ell^+\ell^-}$) of $\mathcal{B}(\mathrm{Z}\to\mathrm{J}/\psi\ell^+\ell^-)$ relative to $\mathcal{B}(\mathrm{Z}\to\mu^+\mu^-\mu^+\mu^-)$ is measured in the fiducial phase-space of the CMS detector.

\section{Selections}

This analysis follows closely the $\mathrm{Z}\to\ell^+\ell^-\ell^+\ell^-$ analysis from CMS~\cite{CMSZ4l}. An ensemble of high-$p_\mathrm{T}$ single, dilepton and lower-$p_\mathrm{T}$ three-lepton triggers is used. The selection criteria for the signal and reference channels are summarised in table~\ref{tab:phasespaceMCtruth}.

\begin{table}[h!]
\begin{center}
\caption{\label{tab:phasespaceMCtruth} Selection criteria for signal and reference channels. Here, $\ell$ refers to a prompt muon or electron from the signal decay, or to either of the two leading muons from the reference decay.  Subscripts give the lepton $p_\mathrm{T}$ rank in decreasing order.}
\begin{tabular}{l}
Selection requirement\\
\hline\\[-2.0ex]
Signal (reference) sample: $0\,(40)<m_{\ell^{+}\ell^{-}}<80\,\mathrm{GeV}$\\
Reference sample:  $4<m_{\mu^\pm_3\mu^\mp_4}<80\,\mathrm{GeV}$\\
$|m_{\mu^+\mu^-\ell^+\ell^-}-91.2\,\mathrm{GeV}|<25\,\mathrm{GeV}$\\
$|\eta(\mathrm{electrons})|<2.5$,\quad  $|\eta(\mathrm{muons})|<2.4$ \\
$p_\mathrm{T}(\ell_1,\ell_2,\mu_3,\mu_4)>(30,15,3.5,3.5)\,\mathrm{GeV}$\\
Signal sample:  $p_\mathrm{T}^{\mathrm{J}/\psi}>8.5\,\mathrm{GeV}$\\
\end{tabular}
\end{center}
\end{table}

In addition, the four leptons, and the two muons coming from the $\psi$ decay, are required to be fitted to a common vertex with a $\chi^2$ probability greater than $5\%$. Finally, all leptons are required to be isolated and have an impact parameter significance $\mathrm{IP}/\sigma_\mathrm{IP}<4$, where $\mathrm{IP}$ is the distance of closest approach of the lepton track to the event vertex and $\sigma_\mathrm{IP}$ is its associated uncertainty.

\begin{figure}[h!]
\centering
{\includegraphics[width=0.4\linewidth]{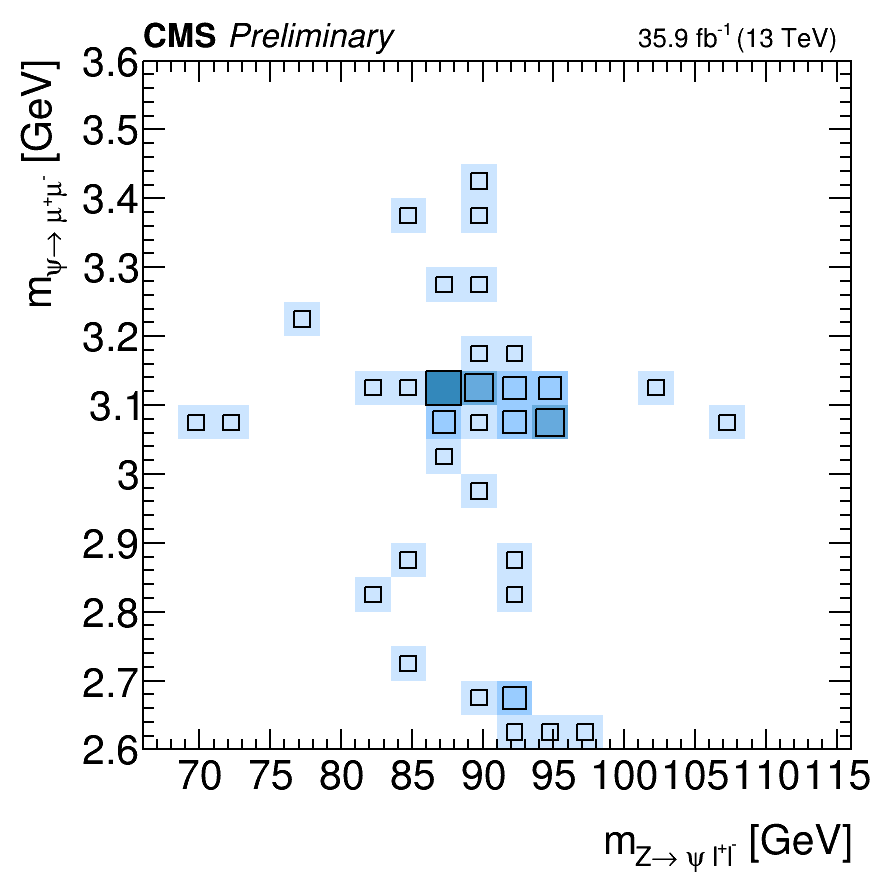}}
\caption{Distribution of invariant masses $m_{\mu^+\mu^-}$ vs. $m_{\mathrm{Z}\to\psi\ell^+\ell^-}$ for selected candidates$^{5}$.}
\label{fig:AllEventsCombined}
\end{figure}

After applying the selection criteria, $29$ $\mathrm{Z}\to\psi\mu^+\mu^-$ and $18$ $\mathrm{Z}\to\psi e^+e^-$ candidate events are found. The $\psi\to\mu^+\mu^-$ and $\mathrm{Z\to\psi\ell^+\ell^-}$ invariant mass distributions of these events are shown in figure~\ref{fig:AllEventsCombined}, where four contributions can be seen. First, the signal $\mathrm{Z}\to\psi\ell^+\ell^-$ decay populating the centre of the two-dimensional (2D) plot, second the $\mathrm{Z}\to\mu^+\mu^-\ell^+\ell^-$ decay, where the $m_{\mu^+\mu^-}$ is non-resonant, third the $\psi\to\mu^+\mu^-$ decay with two additional non-resonant leptons and fourth, two muons and two leptons coming from combinatorial background.

\section{Observation of the $\mathrm{Z}\to\psi\ell^+\ell^-$ decay}

\begin{figure}[h!]
\centering
{\includegraphics[width=0.4\textwidth]{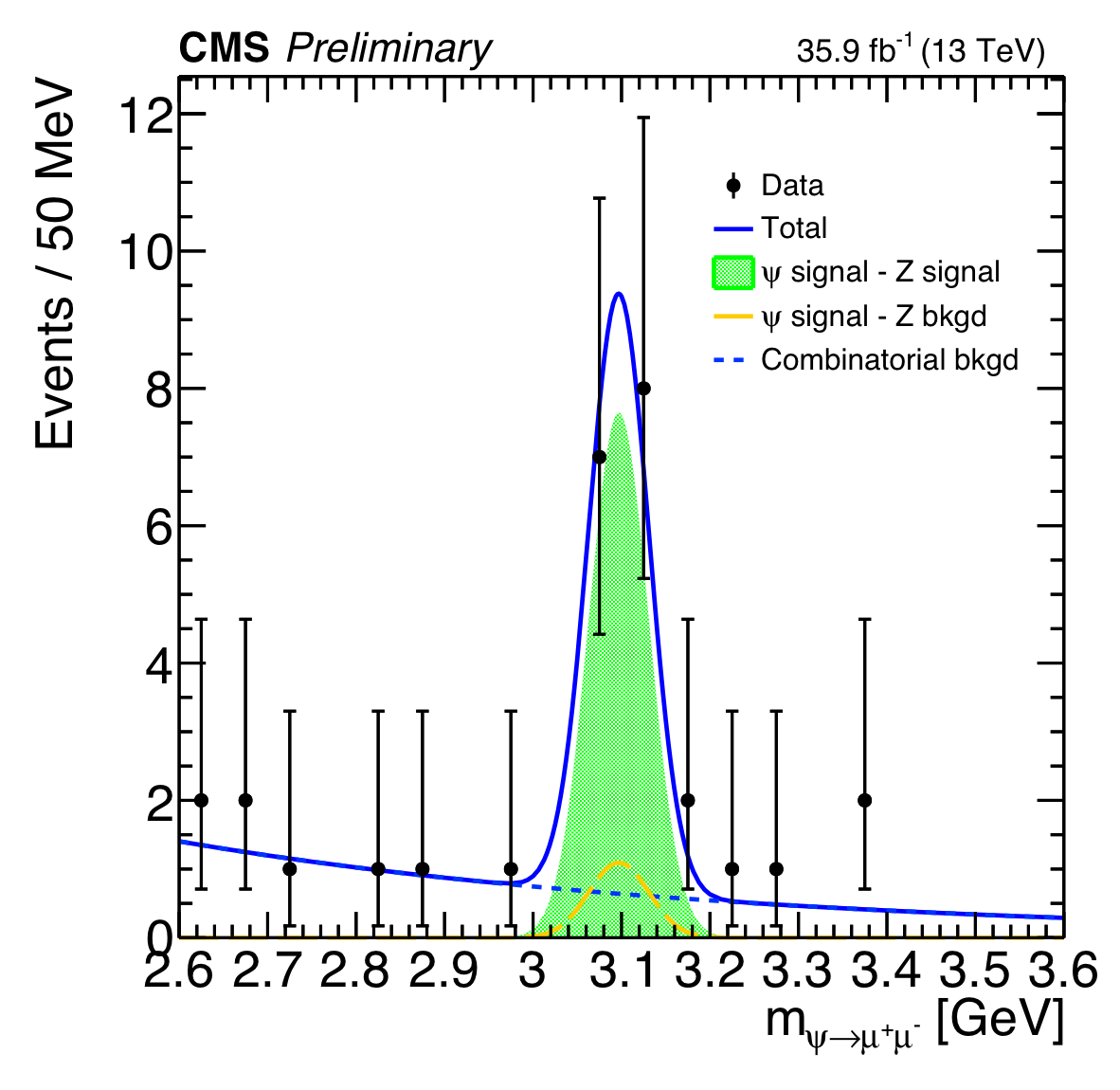}}
{\includegraphics[width=0.4\textwidth]{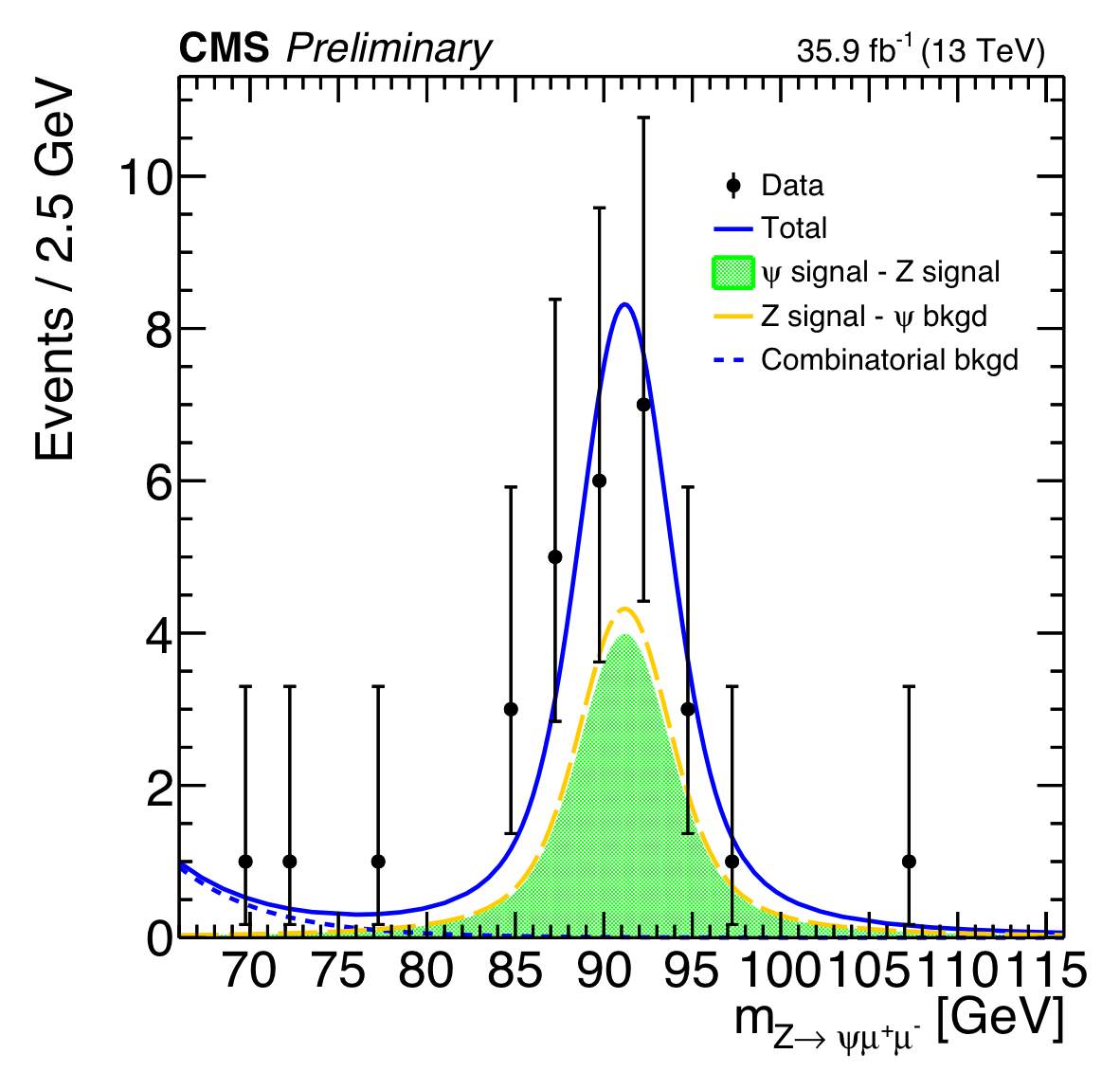}}\\
{\includegraphics[width=0.4\textwidth]{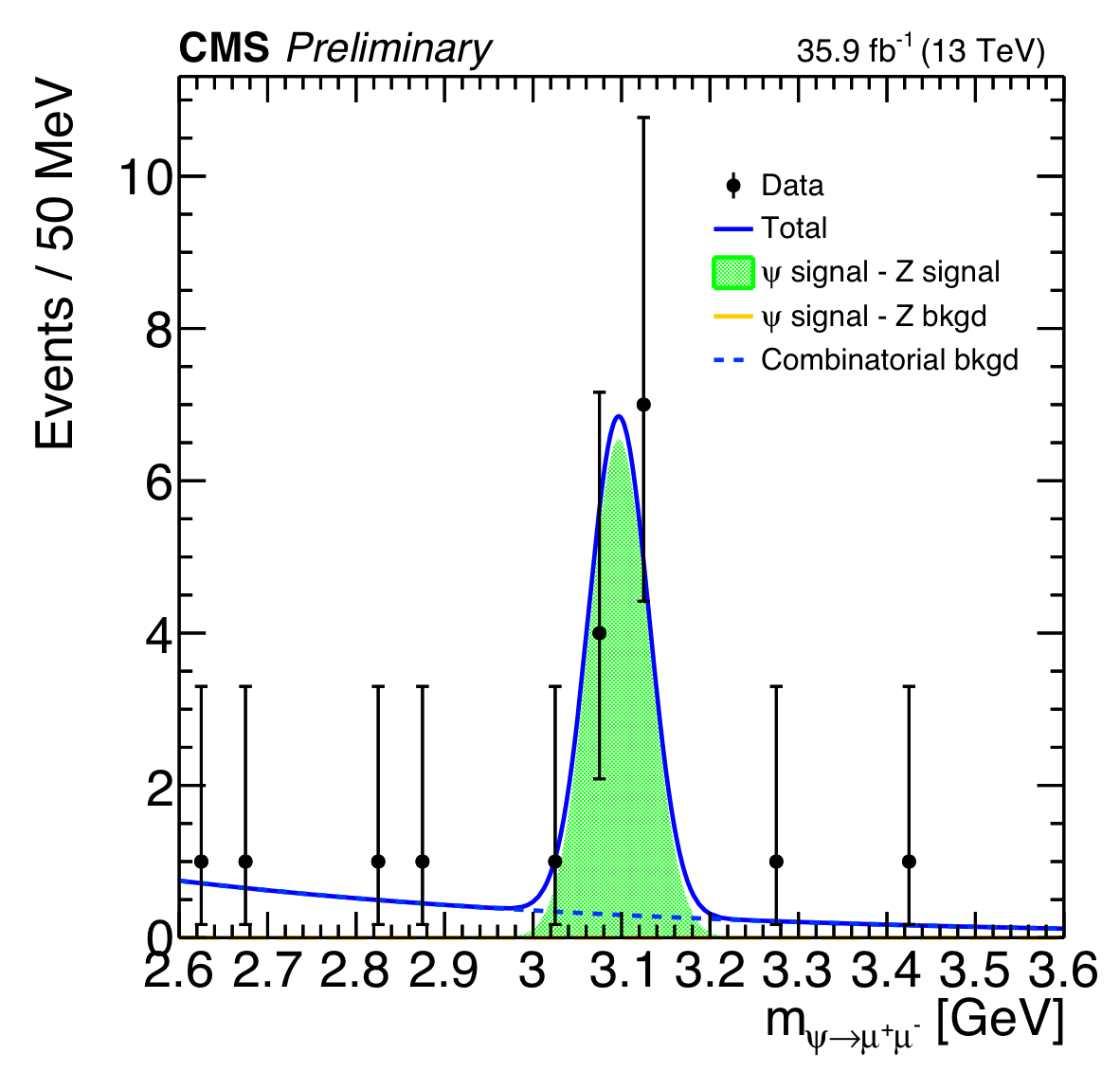}}
{\includegraphics[width=0.4\textwidth]{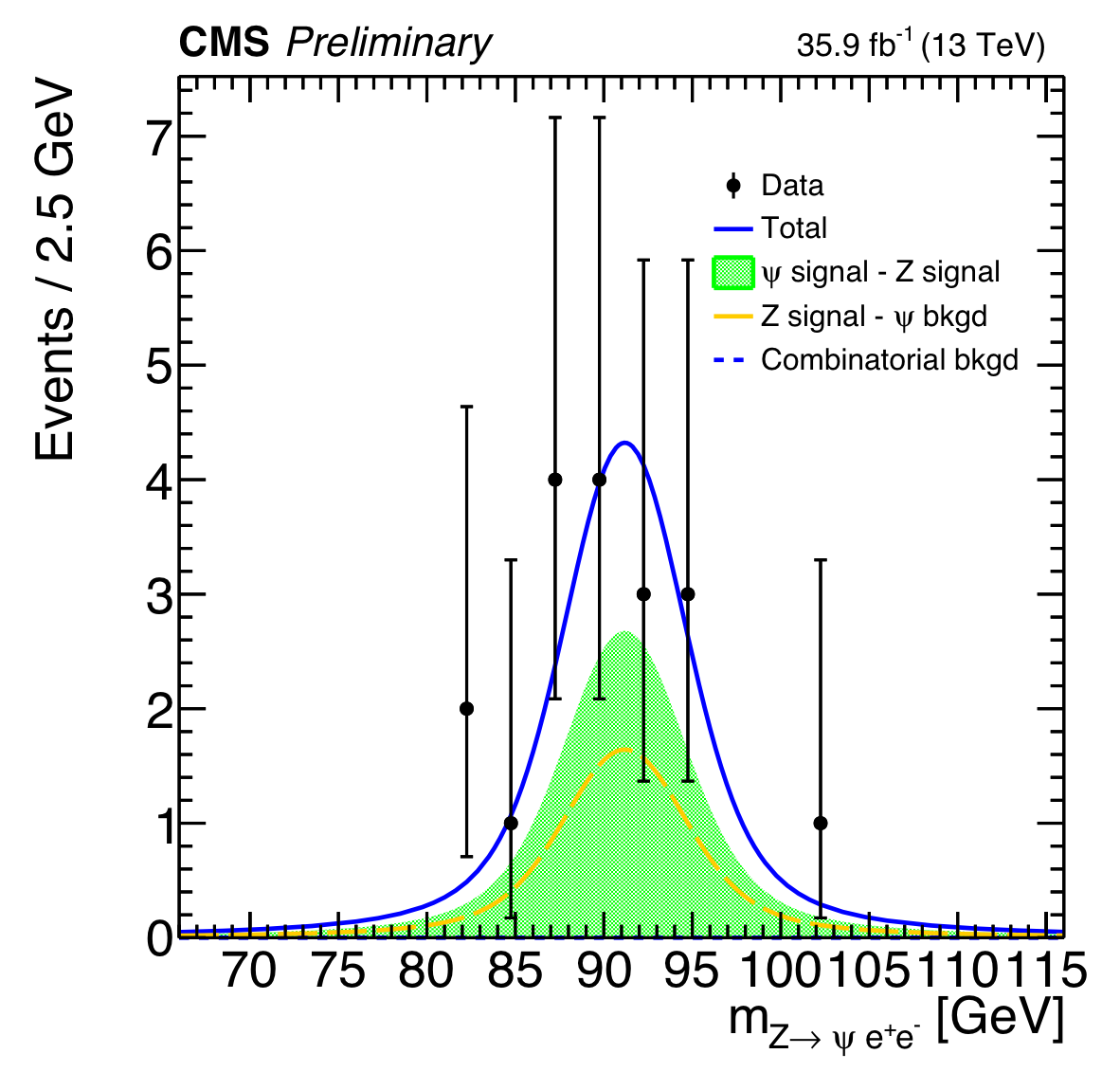}}
\caption{Invariant mass distributions for the $\psi$ muon pairs (left) and for $\mathrm{Z}\to\psi\ell^+\ell^-$ (right), for $\mathrm{Z}\to\psi\mu^+\mu^-$ (upper) and  $\mathrm{Z}\to\psi \mathrm{e}^+\mathrm{e}^-$ (lower) candidates. Data are represented with filled circles, the solid blue line is the overall fit to the data, the green filled region corresponds to the signal yield, while the dashed blue lines are the $\psi$ signal from the $\mathrm{Z}$ background (left) and the $\mathrm{Z}$ signal extracted from the $\psi$ background (right)$^{5}$.}
\label{fig:AllEventsZpsi}
\end{figure}

An extended 2D unbinned maximum likelihood fit is used to determine the signal yield in the $m_{\mathrm{Z}\to\psi\ell^+\ell^-}$ and $m_{\psi\to\mu^+\mu^-}$ variables. The one-dimensional projections of the fit are shown in figure~\ref{fig:AllEventsZpsi}. The number of signal events found are $N_{\mathrm{Z}\to\psi\mu^+\mu^-}=13.0\pm 3.9$ and $N_{\mathrm{Z}\to\psi e^+e^-}=11.2\pm 3.4$.

The signal significance for the $\mathrm{Z}\to\psi\mu^+\mu^-$ and $\mathrm{Z}\to\psi e^+e^-$ decay channels is $4.0\,\sigma$ and $4.3\,\sigma$, respectively. The combined significance is $5.7\,\sigma$.


For the selection of the $\mathrm{Z}\to\mu^+\mu^-\mu^+\mu^-$ reference channel, the same selection criteria as for $\mathrm{Z\to\psi(\to\mu^+\mu^-)\mu^+\mu^-}$ are applied, with the exception of the vertexing of the $\psi$ muons and the $p_\mathrm{T}^{\psi}$ requirement. The $\mathrm{Z}\to\mu^+\mu^-\mu^+\mu^-$ signal is extracted using the same parameterisation used for $\mathrm{Z}$ modelling on the $\mathrm{Z\to\psi\mu^+\mu^-}$ signal model, which results in a signal yield of $250\pm 20$ events.

\section{Fiducial branching fraction measurement}

The fiducial branching fraction of $\mathrm{Z}\to\mathrm{J}/\psi\,\ell^+\ell^-$ relative to the $\mathrm{Z}\to\mu^+\mu^-\mu^+\mu^-$ decay mode is calculated using the formula:

$$
\mathcal{R}_{\mathrm{J}/\psi\ell^+\ell^-} \equiv \frac{\mathcal{B}(\mathrm{Z}\to\mathrm{J}/\psi\ell^+\ell^-)}{\mathcal{B}(\mathrm{Z}\to\mu^+\mu^-\mu^+\mu^-)} =
\sum_{\ell}{\left(\frac{1}{2}\frac{N^\ell_{\mathrm{Z}\to\mathrm{J}/\psi\ell^+\ell^-}}{\epsilon^\ell_{\mathrm{Z}\to\mathrm{J}/\psi\ell^+\ell^-}}\right)} \frac{\epsilon_{\mathrm{Z}\to\mu^+\mu^-\mu^+\mu^-}}{N_{\mathrm{Z}\to\mu^+\mu^-\mu^+\mu^-}}\frac{1}{\mathcal{B}(\mathrm{J}/\psi\to\mu^+\mu^-)}, \ell=\mu,\mathrm{e},
\label{eq:brformula}
$$

where $N^\ell_{\mathrm{Z}\to\mathrm{J}/\psi\ell^+\ell^-}$ is the number of $\mathrm{Z}\to\mathrm{J}/\psi\ell^+\ell^-$ events, after removing $\psi\mathrm{(2S)}\to\mathrm{J}/\psi\,X$ contributions. Using the branching fraction ratio $\mathcal{B}[\mathrm{Z}\to\mathrm{J}/\psi\ell^+\ell^-]/\mathcal{B}[\mathrm{Z}\to\psi(\mathrm{2S})\ell^+\ell^-]$ from Ref~\cite{theorypaper2}, $1.9$ events from $N_{\mathrm{Z}\to\psi\mu^+\mu^-}$ and $1.7$ events from $N_{\mathrm{Z}\to\psi e^+e^-}$ are subtracted, leaving $11.1$ $N_{\mathrm{Z}\to\mathrm{J}/\psi\mu^+\mu^-}$ and $9.5$ $N_{\mathrm{Z}\to\mathrm{J}/\psi\mu^+\mu^-}$ signal events. The efficiencies are evaluated using MC in the fiducial region defined in table~\ref{tab:phasespaceMCtruth} and are found to be $\epsilon_{\mathrm{Z}\to\mu^+\mu^-\mu^+\mu^-}=81.1\%$, $\epsilon_{\mathrm{Z}\to\mathrm{J}/\psi\mu^+\mu^-}=80.8\%$ and $\epsilon_{\mathrm{Z}\to\mathrm{J}/\psi e^+e^-}=79.6\%$. Using these values, the branching fraction ratio is measured to be: $\mathcal{R}_{\mathrm{J}/\psi\ell^+\ell^-}=0.70\pm 0.18\, \mathrm{(stat)} \pm 0.05\, \mathrm{(syst)}$.

This result assumes that the $\mathrm{J}/\psi$ mesons are produced unpolarised. Comparing the results obtained using the unpolarised and the longitudinally polarised and the three transversely polarised scenarios, `transverse 0', `transverse $-$' and `transverse +'~\cite{polarisation}, a variation from $-24$ to $+22\%$ is observed.

Extrapolating from the fiducial region to the full space and assuming that the extrapolations of the signal and reference channels cancel in the ratio a qualitative estimate of the $\mathcal{B}(\mathrm{Z}\to\mathrm{J}/\psi\,\ell^+\ell^-)$ can be extracted. Using $\mathcal{B}(\mathrm{Z}\to\mu^+\mu^-\mu^+\mu^-)=(1.20\pm 0.08)\times 10^{-6}$ for $m_{\mu^+\mu^-}>4\,\mathrm{GeV}$~\cite{CMSZ4l}, $\mathcal{B}(\mathrm{Z}\to\mathrm{J}/\psi\,\ell^+\ell^-)\approx 8\times 10^{-7}$, which is consistent with $(6.7\pm 0.7)\times 10^{-7}$ and $7.7\times 10^{-7}$ as calculated in Ref.~\cite{theorypaper1} and Ref.~\cite{theorypaper2}, respectively.


\section*{Acknowledgments}
The author would like to acknowledge DOE (USA) for the enduring support for the construction and operation of the LHC and the CMS detector.



\section*{References}

\begin{thebibliography}{99}
\bibitem{LHCmachine} L. Evans and P. Bryant, LHC Machine, \Journal{JINST}{3}{S08001}{2008}

\bibitem{theorypaper1} L. Bergstr{\"o}m and R. W. Robinett, $\mathrm{Z}$ decay into a vector meson and a lepton pair, \Journal{\PLB}{245}{249}{1990}

\bibitem{theorypaper2} S. Fleming, Electromagnetic production of quarkonium in $\mathrm{Z}$ decay, arXiv:hep-ph/9304270, \Journal{\PRD}{48}{1914}{1993} 

\bibitem{theorypaper3} S. Fleming, $\mathrm{J}/\psi$ production from electromagnetic fragmentation in $\mathrm{Z}$ decay, arXiv:hep-ph/9403396, \Journal{\PRD}{50}{5808}{1994}

\bibitem{ATLAS1} ATLAS Collaboration, Search for Higgs and $\mathrm{Z}$ boson decays to $\mathrm{J}/\psi\gamma$ and $\Upsilon\mathrm{(nS)}\gamma$ with the ATLAS detector, arXiv:1501.03276, \Journal{PRL}{114}{12}{2015}

\bibitem{ATLAS2} ATLAS Collaboration, Search for exclusive Higgs and $\mathrm{Z}$ boson decays to  $\phi\gamma$ and $\rho\gamma$ with the ATLAS detector, arXiv:1712.02758 (2017)

\bibitem{mypaper} CMS Collaboration, {Observation of the decay $\mathrm{Z}\to\psi\,\ell^+\ell^-$ in $\mathrm{pp}$ collisions at $\sqrt{s}=13\,\mathrm{TeV}$}, {CMS-PAS-BPH-16-001} \url{https://cds.cern.ch/record/2308113/} (2018)

\bibitem{CMS} CMS Collaboration, The CMS experiment at the CERN LHC, \Journal{JINST}{3}{S08004}{2008}

\bibitem{CMSZ4l} CMS Collaboration, Measurements of the $\mathrm {p}\mathrm {p}\rightarrow\mathrm{Z}\mathrm{Z}$ production cross section and the $\mathrm{Z}\rightarrow 4\ell $ branching fraction, and constraints on anomalous triple gauge couplings at $\sqrt{s} = 13\,\mathrm{TeV} $, arXiv:1709.08601, Eur. Phys. J. C {\bf{78}}, 165 (2018)

\bibitem{polarisation} P. Faccioli, C. Louren\c{c}o, J. Seixas and H. K. Wohri, Towards the experimental clarification of quarkonium polarization, arXiv:1006.2738, Eur. Phys. J. C {\bf{69}}, 657 (2010)

\end{thebibliography}
\end{document}